# Molecular Discrimination Using Core-Shell MOF-Coated SAW Devices

Balasubramanian Srinivasan
*Precision Electrochemical Nanosensors*
*Tyndall National Institute*
Cork, Ireland
bala.srinivasan@tyndall.ie

Aswin Shibu
*Precision Electrochemical Nanosensors*
*Tyndall National Institute*
Cork, Ireland
aswin.shibu@tyndall.ie

Veda Sandeep Nagaraja
*Strategic Development*
*Tyndall National Institute*
Cork, Ireland
veda.sandeep@tyndall.ie

Alan O'Riordan
*Precision Electrochemical Nanosensors*
*Tyndall National Institute*
Cork, Ireland
alan.oriordan@tyndall.ie

***Abstract*— A porous sensing layer influences more than mass loading in a surface acoustic wave (SAW) device: molecular adsorption also perturbs its local electrical environment. We investigate this coupling using 128° YX lithium niobate devices coated with ZIF-8 or core–shell ZIF-8@ZIF-67 metal–organic frameworks. The transmission coefficient was monitored at 301 MHz during methanol and acetone exposure (0.5–20 ppm). Changes in linear transmission magnitude followed sublinear power laws. For ZIF-8, fitted exponents were 0.254 for methanol and 0.058 for acetone; introducing the Co-based ZIF-67 shell reversed their ordering to 0.089 and 0.145, respectively. The core–shell coating increased the acetone response by approximately 1.8-fold at 0.5 ppm and 2.5-fold at 20 ppm, with a 4.8-fold enhancement in fitted local concentration sensitivity at 1 ppm. These results show that core–shell architecture modifies both response amplitude and molecule-dependent concentration scaling. Heterogeneous adsorption and altered electrical loading, potentially involving dielectric polarization and interfacial charge redistribution, offer a physical interpretation without directly establishing net molecular charge transfer experimentally. Concentration-response scaling thus complements signal magnitude as an electrical signature for molecular discrimination and real-time environmental monitoring.**

***Keywords—Surface acoustic wave, MOFs, gas sensing, power-law response, Environmental monitoring.***

## I. Introduction

When a molecule enters a porous material, it does more than add mass. Its interaction with the surrounding framework can change the local electrostatic environment, polarization, and charge distribution. These changes provide opportunities for chemical sensing, but distinguishing molecules that produce similar electrical responses remains challenging. Metal–organic frameworks (MOFs) are particularly attractive because their adsorption properties can be influenced by the choice of metal centers, organic linkers, and pore environments [1]. However, high adsorption capacity alone does not guarantee chemical discrimination, especially when different vapors produce overlapping sensing signals [2]. An interesting possibility emerges when two MOFs with similar framework chemistry but different metal centers are combined. ZIF-8 and ZIF-67 share the same 2-methylimidazolate linker but contain $Zn^{2+}$ and $Co^{2+}$ centers, respectively. Incorporating ZIF-67 as a shell around ZIF-8 introduces a second metal-node environment and an additional interface while retaining closely related framework chemistry. *The important question is therefore not simply whether the core–shell MOF produces a larger response, but whether introducing a second metal-centered framework changes how the sensor responds to different molecules.*

Surface acoustic wave (SAW) devices offer a sensitive platform for investigating such interactions. Operating at radio frequencies, they generate rapidly oscillating electromechanical fields concentrated near the functionalized surface. At 300 MHz, the oscillation period is approximately 3.3 ns, providing high-frequency interrogation of adsorption-induced perturbations in the local surface environment. MOF-coated SAW devices have demonstrated gas detection at ppm concentrations [3], including ZIF-functionalized devices for methane and carbon dioxide [4]. Recent advances in functional coatings have further improved acetone sensing performance [5]. SAW devices are commonly evaluated through figures of merit such as quality factor, insertion loss, and electromechanical coupling, while gas-sensing performance is assessed through sensitivity, response magnitude, and detection limit. Although essential, these quantities do not necessarily reveal molecule-specific interactions: similar response magnitudes may conceal distinctly different concentration-dependent behaviour.

This distinction is particularly important because SAW devices respond to both mechanical and electrical perturbations. Electrical loading and acoustoelectric coupling can influence SAW propagation [6], while molecular adsorption within a MOF may additionally modify dielectric polarization, effective capacitance, and interfacial charge distribution [7]. These changes can perturb the impedance and RF transmission of the interdigital structure. Of particular interest is the electrical response near antiresonance, where capacitive and motional contributions can partially cancel, making the device sensitive to changes in electrical loading. The ZIF-8@ZIF-67 interface may further influence this electrical response [8]. Earlier electrical-resonance studies have demonstrated how adsorption-induced capacitance changes can be related to equivalent charge modulation [9].

Here, we investigate whether introducing a Co-based ZIF-67 shell modifies the VOC-sensing characteristics of Zn-based ZIF-8 using nominally 300 MHz SAW devices. Methanol and acetone are selected as probe molecules to compare the two coating architectures. Rather than tracking resonance-frequency shifts, we monitor the transmission parameter $S_{21}$ at 301 MHz, near a pronounced antiresonance-associated transmission minimum. Beyond conventional response magnitude, we investigate the fractional power-law exponent describing the concentration dependence of the electrical response. A simple Freundlich-type adsorption description provides a physical basis for considering the observed sublinear behaviour without equating the response exponent directly with an adsorption-isotherm parameter. Together, these measurements explore whether core–shell MOF architecture introduces additional electrical response signatures for distinguishing molecular adsorption.

## II. EXPERIMENTAL METHODS

### *A. MOF Synthesis*

ZIF-8 was synthesized by magnetic stirring at room temperature. $Zn(NO_3)_2{\cdot}6H_2O$ (0.2933 g) and 2-methylimidazole (HMIM; 0.6489 g) were separately dissolved in 20 mL of methanol. The zinc nitrate solution was added to the stirred HMIM solution and mixed for 24 h in a closed container. The resulting white precipitate was centrifuged at 6000 rpm for 15 min, washed three times with 40 mL of methanol, and redispersed in 40 mL of methanol for storage in a sealed container. For ZIF-8@ZIF-67 synthesis, ZIF-8 (0.25 g) was dispersed in 50 mL of methanol. $Co(NO_3)_2{\cdot}6H_2O$ (2.91 g, 0.01 mol) and 2-methylimidazole (3.08 g, 0.0375 mol) were separately dissolved in 50 mL of methanol each. The cobalt nitrate solution was added to the ZIF-8 dispersion, followed by the 2-methylimidazole solution. After stirring at room temperature for 24 h, the resulting light-purple precipitate was centrifuged, washed repeatedly with methanol, and redispersed in methanol for storage in a sealed container.

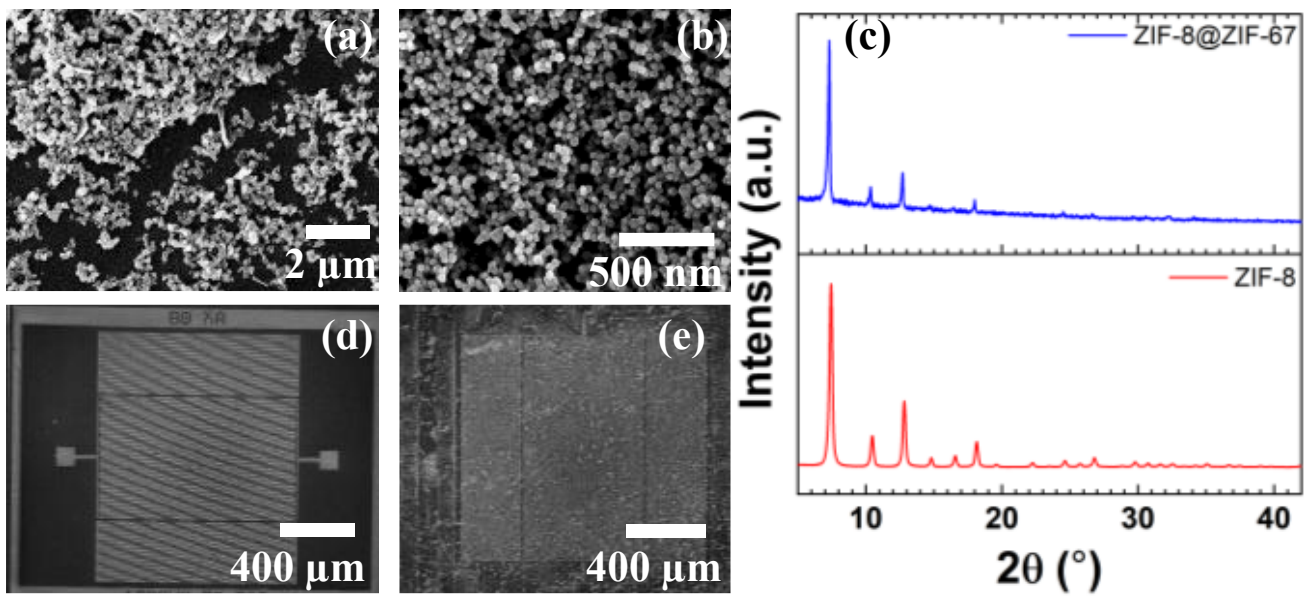

Fig. 1. Morphological and structural characterization of the MOF sensing materials and SAW devices: SEM images of (a) ZIF-8@ZIF-67 core–shell MOF and (b) ZIF-8; (c) XRD patterns of ZIF-8 and ZIF-8@ZIF-67; and SEM images of the SAW device (d) before and (e) after MOF coating.

### *B. SAW Coating and Characterization*

Both MOFs were deposited onto separate SAW devices by controlled drop-casting from methanolic dispersions at 40 °C. A glass slide positioned above each device formed a sandwiched configuration to suppress coffee-ring formation and promote uniform deposition. The coated devices were vacuum-dried overnight at 50 °C and stored in a vacuum desiccator. Scanning electron microscopy (SEM) was used to examine particle morphology and coating distribution, while X-ray diffraction (XRD) characterized the crystalline phases (Fig. 1).

### *C. Gas exposure and RF readout*

The experimental setup is illustrated in Fig. 2a. Dry nitrogen ($N_2$) was supplied through two mass-flow-controlled lines, providing carrier and dilution flows. Methanol or acetone was introduced using a syringe pump into the vapor-generation line and mixed with the controlled $N_2$ flow to obtain the desired gas concentrations. The resulting vapor mixture was delivered to a temperature-controlled sensing chamber, with the outlet directed to a fume hood. The total gas flow rate was maintained at 100 sccm throughout the experiments. The chamber temperature was maintained at 25 °C under ambient pressure and humidity conditions. VOC concentrations ranged from 0.5 to 20 ppm. The sensing device comprised a single two-terminal interdigital transducer fabricated on 128° YX $LiNbO_3$, with a nominal operating frequency of 300 MHz. Each terminal was connected to a separate calibrated port of a vector network analyzer (VNA). The electrical transmission parameter $S_{21}$ was monitored at 301 MHz during alternating $N_2$ baseline and VOC exposure periods. The VNA excitation power was maintained at −5 dBm for all measurements, corresponding to an incident RF power of approximately 0.316 mW in the 50 Ω reference system. The measured $S_{21}$ values in dB were converted to linear transmission magnitudes. The percentage response relative to the preceding $N_2$ baseline was calculated as,

$$R(C) = 100\left|10^{\frac{s(C)-s_0}{20}} - 1\right| \qquad (1)$$

where $s(C)$ and $s_0$ denote the gas-exposed and preceding $N_2$ baseline $S_{21}$ readings in dB, respectively. $R(C)$ represents the absolute percentage change in linear transmission magnitude. The response magnitude was used to evaluate the concentration-dependent sensing characteristics.

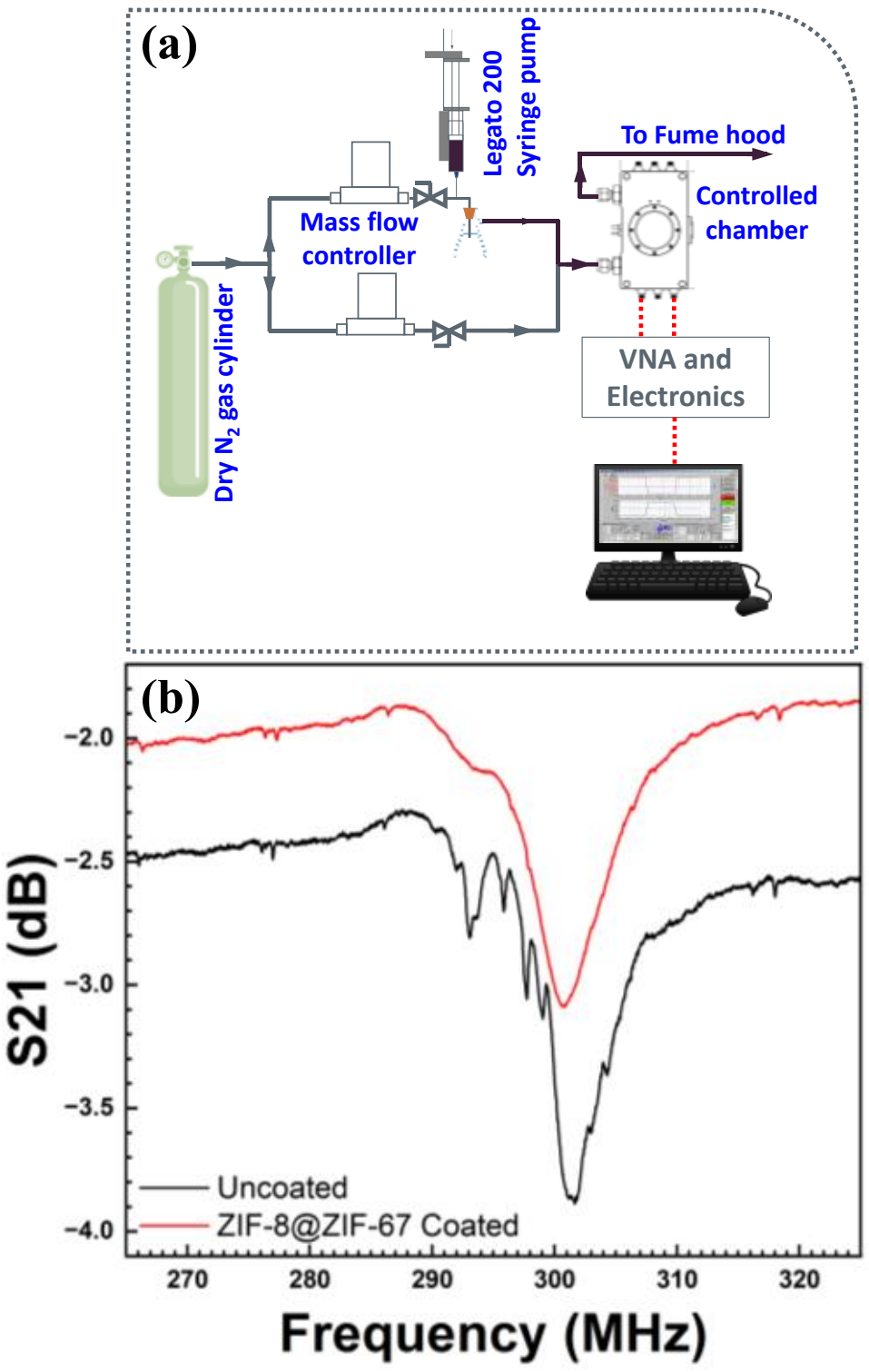

Fig. 2. (a) Experimental VOC sensing setup with controlled vapor delivery and two-port VNA readout. (b) $S_{21}$ spectra of uncoated and ZIF-8@ZIF-67-coated SAW devices, showing a transmission minimum near 301 MHz selected for gas sensing.

### *D. RF Characterization*

Figure 2(b) compares the RF transmission spectra of the uncoated and ZIF-8@ZIF-67-coated devices. Analysis of uncoated-device RF measurements yielded an apparent effective electromechanical coupling coefficient ($k_{eff}^2$) of 6.52 ± 0.46% and a conductance-linewidth Q-factor of 42.1 ± 2.1. The ZIF-8@ZIF-67-coated device exhibited a transmission minimum at 300.66 MHz and an apparent conductance-linewidth Q-factor of approximately 40 ± 1.5. The comparable apparent Q values indicate that well-defined RF features are retained following coating, although their different conductance characteristics preclude a direct comparison of intrinsic acoustic damping. The pronounced transmission minimum near 301 MHz was selected for subsequent VOC sensing experiments.

## III. RESULTS AND DISCUSSION

### *A. Concentration-dependent electrical response*

Figure 3 compares the responses of both coatings toward methanol and acetone over 0.5–20 ppm. The relative changes in linear transmission magnitude were fitted using

$$R(C) = A\left(\frac{C}{C_{\mathrm{ref}}}\right)^{\beta} \qquad (2)$$

$A$ is the fitted response at the reference concentration, $C_{\mathrm{ref}}$is the reference concentration, chosen as 1 ppm and $\beta$ is the concentration-response exponent. The corresponding local sensitivity is

$$S(C) = \frac{dR}{dC} = \frac{\beta R(C)}{C} \qquad (3)$$

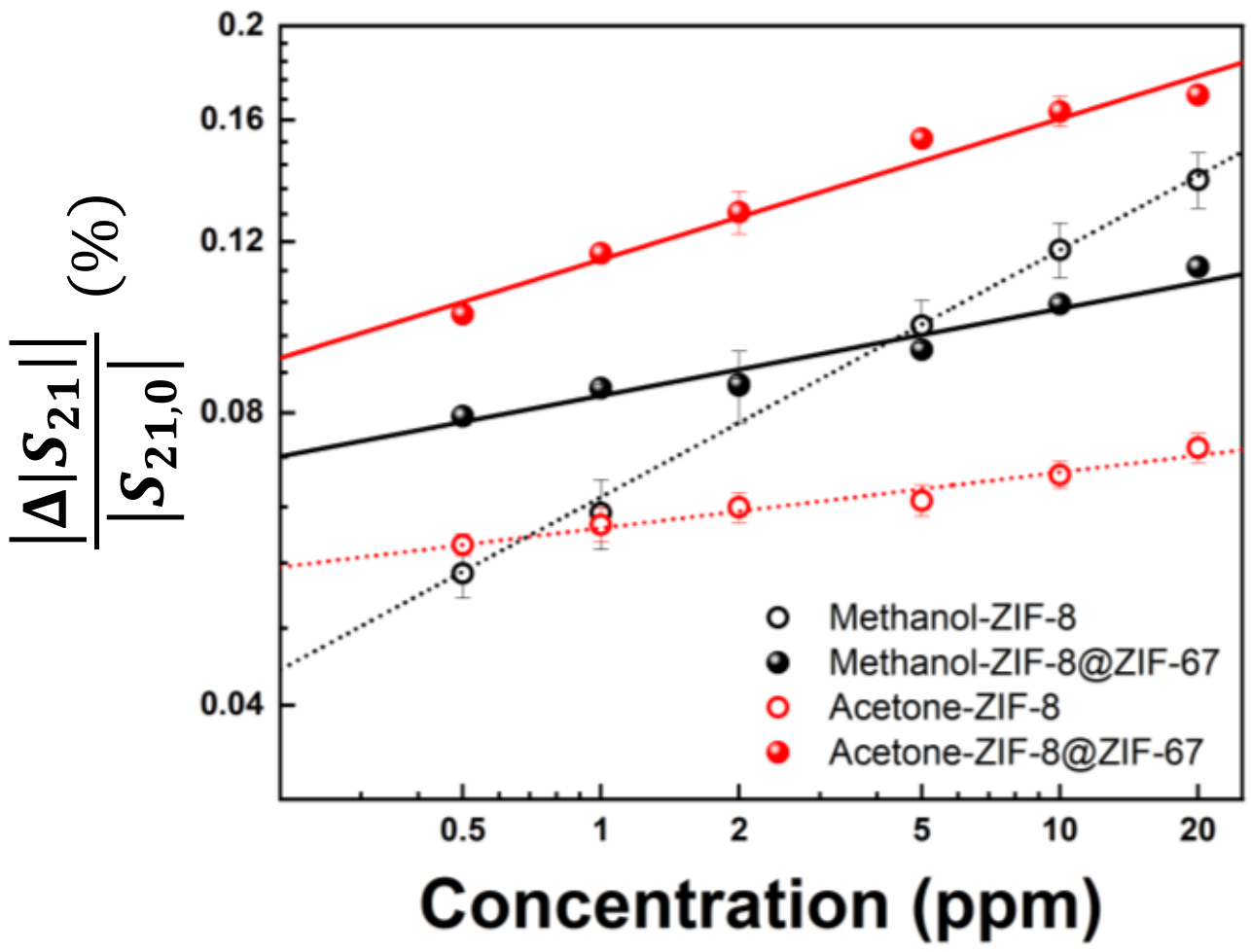

Fig. 3. Concentration-dependent relative shift in linear $S_{21}$ magnitude for methanol (black) and acetone (red) on ZIF-8 (open symbols) and ZIF-8@ZIF-67 (filled symbols) over 0.5–20 ppm. Lines represent empirical power-law fits.

TABLE I. POWER-LAW PARAMETERS AND FITTED LOCAL SENSITIVITIES AT 1 PPM

| Coating / VOC | A (%) | β | *S* (1 ppm) (%/ppm) |
|---|---|---|---|
| ZIF-8 / Methanol | 0.066 | 0.25 | 0.017 |
| ZIF-8 / Acetone | 0.061 | 0.058 | 0.0035 |
| ZIF-8@ZIF-67 / Methanol | 0.083 | 0.090 | 0.007 |
| ZIF-8@ZIF-67 / Acetone | 0.115 | 0.15 | 0.017 |

For ZIF-8, methanol exhibits a substantially larger exponent than acetone, indicating stronger concentration-dependent scaling. Interestingly, introducing the Co-based ZIF-67 shell reverses this relationship: the methanol exponent decreases from 0.25 to 0.090, whereas the acetone exponent increases from 0.058 to 0.15. This reversal indicates that the core–shell architecture modifies the relative molecular-response characteristics rather than uniformly enhancing the sensing signal. The response magnitude reveals a complementary distinction. For acetone, the core–shell coating produces approximately 1.8- and 2.5-fold higher fitted responses at 0.5 and 20 ppm, respectively, with a 4.8-fold increase in local sensitivity at 1 ppm. In contrast, the core–shell methanol response is higher at low concentrations but becomes smaller than that of ZIF-8 above approximately 4.3 ppm. The ZIF-67 shell therefore preferentially enhances acetone responsiveness while modifying the concentration scaling of methanol.

Understanding this behaviour requires considering how molecules access and interact with the porous frameworks. ZIF-8 consists of approximately 11.6 Å cavities connected by six-membered-ring windows with a nominal crystallographic aperture of 3.4 Å [10]. ZIF-67 shares a closely related sodalite structure; thus, introducing the Co-containing shell does not simply enlarge the pore openings but modifies the metal-node environment and introduces an additional interface. Importantly, a molecule may fit within a cavity yet encounter a substantial barrier at its connecting window. Although methanol is smaller than acetone and would be expected to experience less steric hindrance, ZIF frameworks are not rigid molecular sieves. Rotation of the 2-methylimidazolate linkers can modify the effective aperture, allowing molecular access beyond that predicted by the static crystal structure [10], [11]. Guest accessibility therefore depends on molecular dimensions, framework flexibility, and guest–framework interactions, as demonstrated for alcohols in ZIF-8 [12]. In ZIF-8@ZIF-67, molecules encounter the Co-containing shell, while external surfaces, grain boundaries, and accessible interfaces may provide additional adsorption environments. Consequently, the stronger acetone response does not necessarily indicate faster diffusion or greater molecular uptake; it may instead arise from differences in adsorption affinity, accessible binding environments, or a stronger electrical perturbation per adsorbed molecule. This interpretation can be developed through the Freundlich adsorption relationship,

$$q_{\mathrm{eq}}(p) = K_F\left(\frac{p}{p_{\mathrm{ref}}}\right)^{n} \qquad (4)$$

where $q_{\mathrm{eq}}$ is the equilibrium adsorbed amount, $K_F$ is an adsorption coefficient, and $n$ describes the effective dependence on analyte partial pressure over a specified range. For a heterogeneous adsorption environment, high-affinity sites can contribute disproportionately at low pressure, while additional sites become populated as pressure increases. The resulting uptake may exhibit sublinear behaviour, represented empirically by $0 < n < 1$. The ZIF frameworks provide plausible origins for such heterogeneity, including distinct adsorption positions within the cages, interactions near narrow pore windows, external crystal surfaces, and defects or interfaces. In particular, crystallographic and computational investigations have identified multiple adsorption sites with different interaction energies within ZIF-8 [10]. Introducing a Co-containing shell may modify the distribution of accessible adsorption environments, potentially affecting methanol and acetone differently.

Although the Freundlich isotherm provides a plausible description of the observed sublinear concentration dependence, the fitted exponent $\beta = \frac{d\ln R}{d\ln C}$ need not equal the adsorption exponent ($n$). The measured electrical response reflects not only molecular uptake within heterogeneous adsorption environments but also how adsorption modifies dielectric polarization, interfacial charge distribution, and RF transmission. Furthermore, at the low vapor concentrations investigated here, a Henry-law adsorption regime may also be relevant. The power-law exponent therefore captures a fundamental characteristic of the MOF–SAW sensing interface: how molecular concentration is translated into an electrical response through the combined influence of adsorption and transduction. Importantly, the reversal in methanol and acetone exponents upon introducing the ZIF-67 shell, together with the enhanced acetone sensitivity, suggests that the core–shell architecture modifies how molecule-dependent adsorption within distinct pore and interfacial environments is translated into an electrical response.

### *B. Electrical loading and charge redistribution*

To examine the electrical origin of the VOC response, the real and imaginary components of $S_{21}$ were analyzed at 301 MHz for both coatings. The gas-induced complex transmission perturbation was evaluated relative to the preceding nitrogen baseline as $\delta = \frac{S_{21,g} - S_{21,0}}{S_{21,0}}$, retaining both amplitude and phase information. With a VNA excitation power of −5 dBm, the incident RF power was 0.316 mW, allowing the change in transmitted power to be determined from the measured transmission magnitude. The complex response reveals an important distinction between the two coating architectures. At 20 ppm acetone, the relative transmission-magnitude response increases from 0.075% for ZIF-8 to 0.17% for ZIF-8@ZIF-67, whereas the corresponding phase shifts remain comparable (0.0414° and 0.0434°, respectively). The core–shell architecture therefore produces a substantially greater amplitude response without a corresponding enhancement in the phase perturbation. Methanol exhibits a different concentration-dependent behaviour, consistent with the contrasting response exponents observed in Fig. 3 as shown in Table II.

TABLE II. COMPARISON OF ELECTRICAL TRANSMISSION RESPONSES

| MOF coating | VOC | Relative change in $\Delta\|S_{21}\|$ (%) | Phase shift $\Delta\phi$ (°) |
|---|---|---|---|
| ZIF-8 | Methanol | 0.138 | +0.018 |
| ZIF-8@ZIF-67 | Methanol | 0.122 | +0.021 |
| ZIF-8 | Acetone | 0.075 | +0.041 |
| ZIF-8@ZIF-67 | Acetone | 0.168 | +0.043 |

These observations indicate that the introduction of the Co-containing shell modifies the complex RF transmission response to molecular adsorption. Guest-induced changes in dielectric polarization and interfacial charge distribution may contribute to this behaviour, while electrical losses and acoustic motional loading can also influence transmission near the selected operating frequency. Earlier electrical-resonance studies related capacitance changes to equivalent charge modulation [9]; however, the present single-frequency transmission measurements do not uniquely separate static capacitance from motional contributions. Together, the reversal in concentration-response scaling and the enhanced acetone transmission demonstrate that changing the MOF architecture reshapes the electrical signatures of molecular adsorption, establishing a route toward molecular discrimination that extends beyond conventional response magnitude and sensitivity.

## IV. CONCLUSION

ZIF-8 and ZIF-8@ZIF-67-coated SAW devices exhibit distinct concentration-dependent electrical responses to methanol and acetone. Introducing the Co-based ZIF-67 shell reverses the relative power-law exponents, from 0.254 (methanol) and 0.058 (acetone) for ZIF-8 to 0.089 and 0.145, respectively. The core–shell architecture enhances the fitted local acetone sensitivity by approximately 4.8-fold at 1 ppm and the measured transmission response by 2.24-fold at 20 ppm, while producing comparable phase shifts. These findings are consistent with architecture-dependent adsorption and electrical loading, potentially involving dielectric polarization and interfacial charge redistribution. Together, concentration-dependent response magnitude and complex electrical transmission provide complementary signatures for molecular discrimination through MOF engineering, beyond conventional sensitivity metrics.

## ACKNOWLEDGMENT

This research was supported in part by Science Foundation Ireland (SFI) through CONNECT (13/RC/2077_P2), co-funded by the European Regional Development Fund, and VistaMilk (21/RC/10303_P2), co-funded by the Department of Agriculture, Food and the Marine, Ireland.